\documentclass{article}

\usepackage{arxiv}

\usepackage{amsmath,amssymb,amsfonts}
\usepackage[utf8]{inputenc} 
\usepackage[T1]{fontenc}    
\usepackage{hyperref}       
\usepackage{url}            
\usepackage{booktabs}       
\usepackage{amsfonts}       
\usepackage{nicefrac}       
\usepackage{microtype}      
\usepackage{cleveref}       
\usepackage{lipsum}         
\usepackage{graphicx}
\usepackage{natbib}
\usepackage{doi}

\title{Scaffold-Based Multi-Objective Drug Candidate Optimization}

\usepackage{authblk}

\newbox{\orcid}\sbox{\orcid}{\includegraphics[scale=0.06]{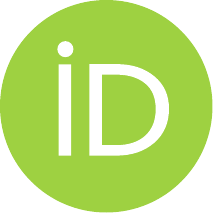}} 

\author[1]{%
	\href{https://orcid.org/0000-0002-5571-7418}{\usebox{\orcid}\hspace{1mm}Agustin~Kruel}%
    }
\author[1]{%
    \href{https://orcid.org/0000-0002-4146-7921}{\usebox{\orcid}\hspace{1mm}Andrew D.~McNaughton}%
    }
\author[1]{%
	\href{https://orcid.org/0000-0001-6713-2129}{\usebox{\orcid}\hspace{1mm}Neeraj~Kumar\thanks{\texttt{neeraj.kumar@pnnl.gov}}}%
    }

\affil[1]{Modeling and Simulation, Pacific Northwest National Laboratory, Richland, WA 99353}

\renewcommand{\shorttitle}{Scaffold MPO}

\begin{document}
\maketitle

\begin{abstract}
In therapeutic design, balancing various physiochemical properties is crucial for molecule development, similar to how Multiparameter Optimization (MPO) evaluates multiple variables to meet a primary goal. While many molecular features can now be predicted using \textit{in silico} methods, aiding early drug development, the vast data generated from high throughput virtual screening challenges the practicality of traditional MPO approaches. Addressing this, we introduce a scaffold focused graph-based Markov chain Monte Carlo framework (ScaMARS) built to generate molecules with optimal properties. This innovative framework is capable of self-training and handling a wider array of properties, sampling different chemical spaces according to the starting scaffold. The benchmark analysis on several properties shows that ScaMARS has a diversity score of 84.6\% and has a much higher success rate of 99.5\% compared to conditional models. The integration of new features into  MPO  significantly enhances its adaptability and effectiveness in therapeutic design, facilitating the discovery of candidates that efficiently optimize multiple properties.
\end{abstract}

\keywords{Drug Design \and Multi-Parameter Optimization \and Neural Network}

\section{Introduction}\label{sec1}

Machine learning (ML) has become increasingly useful for medicinal chemistry, including in the area of drug design. \cite{ML_in_Chemiformatics,criticalAIdrug,AIinDrug} A molecule's structure determines its activity towards biological targets, physiochemical properties, even ease of synthesis. \cite{structure_activity,SynthAccess} It follows that all of these properties must be balanced when designing drug candidates at the risk of becoming toxic or ineffective. Multi-parameter optimization (MPO) outlines ways to rank candidates according to all properties of interest at once, translating changes to a molecule into changes in its ranking. The challenge lies in predicting which changes in molecular structure will contribute to more desirable property values. Some properties with trivial solutions, like cLogP increasing with longer chains, even become non-trivial when in conjunction with other objectives and constraints. \cite{JANUS} ML streamlines the search for an answer with two techniques highlighted in this paper: evolutionary and conditional models.

\begin{figure}[ht]
\centering
\includegraphics[width=0.7\textwidth]{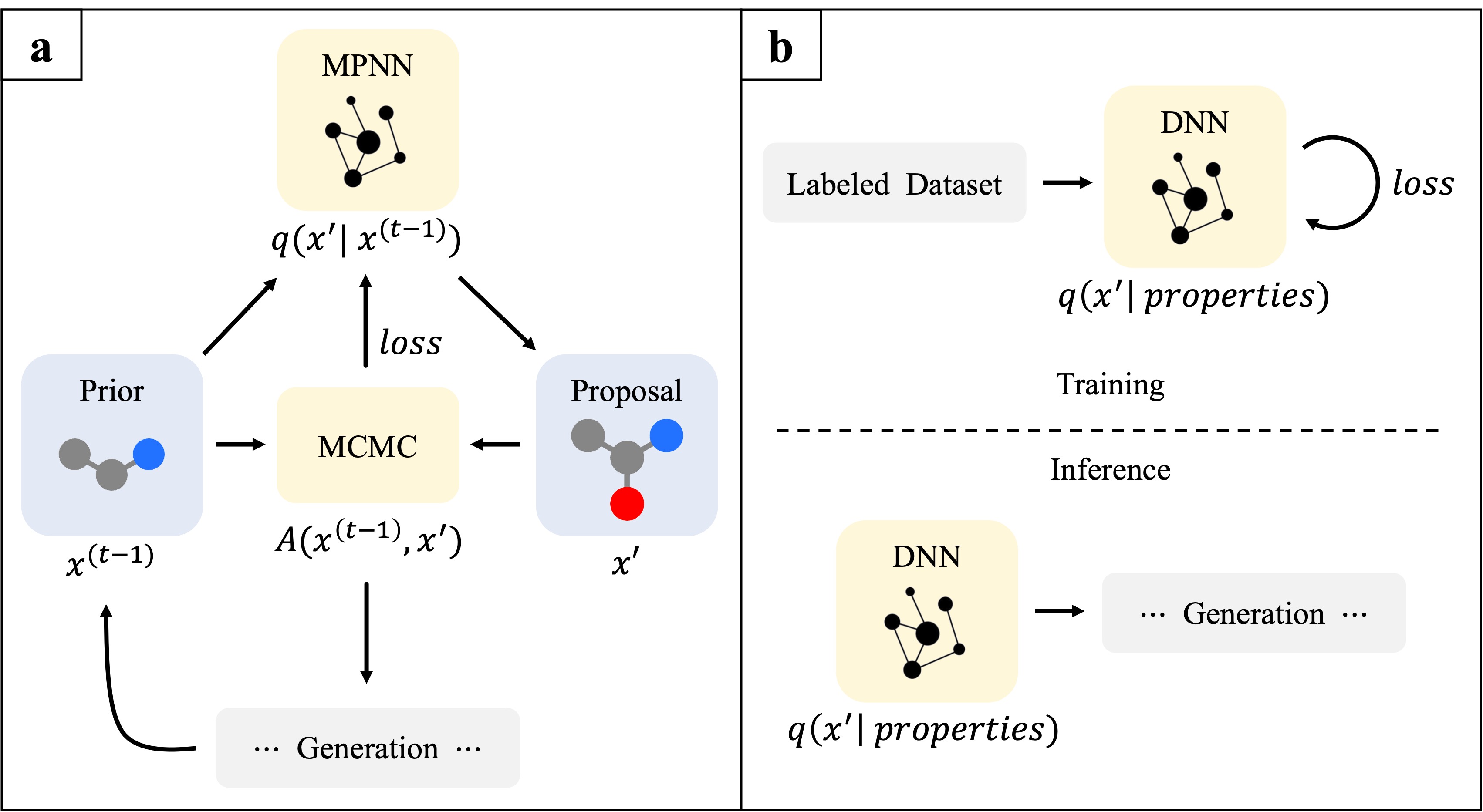}
\caption{(a) ScaMARS workflow for proposing a new molecule generation. First, the initial scaffold or molecule from the previous generation is fed into the MPNN as the 'Prior'. A new, 'Proposal' molecule is then proposed through edits (addition or subtraction of atom groups) by the MPNN. The MPO scores of both Prior and Proposal molecules are used in the annealed MCMC to choose whether the model accepts the proposal. If so, the Proposal is added to the generation and the cycle repeats. If not, the Prior molecule is kept unchanged for the next generation. Once it reaches the desired number of molecules, MPNN loss is calculated on the success of the entire generation to favor beneficial edits. (b) The workflow of a conditional model, specifically molGCT, for comparison.}\label{MARS_arch}
\end{figure}

In evolutionary models, the model navigates through chemical space using iterative changes to a molecule or populations of molecules as movement. The model seeks paths that lead to desirable molecules while avoiding paths that end in toxic or otherwise ineffective solutions. The Graph Neural Network (GNN) model MARS and Genetic Algorithm (GA) JANUS are both such evolutionary models utilizing different architectures. \cite{MARS,JANUS} The compound datasets formed from the collection of all paths in a run will be relatively diverse as well, spanning both optimal and sub-optimal regions of the space as it progresses.

Conditional models, such as the generative chemical Transformer molGCT \cite{molGCT}, instead spend time learning the patterns of a dataset that enable it to generate a new molecule in one step. Given target values for each property, a trained molGCT could output one new molecule that meets such properties. This has common applications in material design and discovery where specific values of properties must be met instead of maximized. \cite{RN6,conditonalSEM,conditionalPhase} What molGCT gains in generation speed and specificity, though, is balanced by slower training and inflexible architecture. Generating based on different choices or a greater number of properties requires retraining the model from scratch (a lengthy, computationally expensive task) and changing its architecture to allow more input nodes.

How the model ranks molecules according to its desirability requires one of two main MPO approaches: Pareto optimization and desirability functions. The former seeks to explore all the balances and trade-offs to the objectives, resulting in all solutions that line the Pareto frontier. These solutions are as likely to excel in only one or two objectives as they are to maximize all at once. Care must also be taken to consider the best function (or ensemble) to quantify this frontier, as chemical space can suffer from complex fronts and discontinuities born from inherent physical limitations or constraints applied to the compounds. \cite{pareto_tech} Given the choice between the two, Pareto optimization becomes infeasible when analyzing the overwhelming quantity of chemical properties. \cite{MPO_Approaches} Without reliable methods to group many disparate properties into a few representative scores, the Pareto frontier increases exponentially with each variable and becomes prohibitively costly to compute. Additionally, data becomes more sparse and increases the proportion deemed non-dominated or stuck in local optima. The compromise of collapsing groups of properties into fewer, manageable scores already overlaps with the second approach, desirability functions. These functions more easily suit ML and the exploration of chemical space by applying mathematical formulae to combine all relevant properties into one number with a Derringer function. \cite{Derringer} This formula can also vary per application according to preferred behavior, such as how mean is calculated, how to normalize each property, or weight values.

In this paper, we adapt a Scaffold-focused MArkov moleculaR Sampling (ScaMARS) model based on the GNN model MARS to be scaffold-focused (ScaMARS). \cite{MARS} ScaMARS is an example of a message passing neural network (MPNN) predicting the probability that one of three actions on a certain node will increase overall properties. It uses Markov chain Monte Carlo (MCMC) to sample proposal molecules after the MPNN adds or subtracts groups of atoms (fragments), treating molecular changes as state transitions. ScaMARS uses a list of 1,000 commonly-occurring fragments extracted from the ChEMBL database. \cite{chembl} Simulated annealing ensures greater compound diversity early (exploration) before settling into globally optimal solutions (exploitation). The MPNN portion learns from each generation it proposes, seeking to improve subsequent generations and eliminating the need for externally annotated data or pretraining.

The importance of explainable AI (XAI) must also be considered. Black-box methods lacking user interpretability also lack the trust needed for important applications such as human safety and life-saving medicine. Some common techniques to explain a trained model post-hoc include feature attribution and subgraph identification, both of which attempt to identify the importance of each input on the output. \cite{DrugXAI} As another technique, decision trees instead mirror the reasoning the model employs and has recently been extended to graph networks. \cite{GAM,DT,DTGNN} The model itself could also be built with interpretability, though, as is the case with ScaMARS and its MCMC layer. At each step, the user can track the journey of the population or each molecule and see which regions of chemical space the model is avoiding or prioritizing. This information can even inform future design.

\section{Methods}\label{sec2}

Figure \ref{MARS_arch} displays a summary of the ScaMARS architecture. ScaMARS's expanded equation allows for many more properties calculated through RDKit \cite{RDKit} as well as alternative desirability functions, while the original MARS paper focused on optimizing two chemical and two ML-predicted biological properties. The user may limit the focus of the optimization run by supplying a list of supported properties or easily add unsupported properties to the equation by adding a python function for it in one file.

\subsection*{Objectives}

Desirable ranges for a molecule's properties depend on the application. The desirability function in ScaMARS is flexible enough to account for user choice in which properties to use, as well as custom formulae for calculation and normalization of properties. This work follows the ranges for assessments of absorption, distribution, metabolism, and excretion (ADME) used in SwissADME \cite{SwissADME} and summarised in Table \ref{objtable}. Whether the function seeks to linearly minimize or maximize the property that falls within the range was determined through trends of each property.
Including the original GSK3$\beta$, JNK3, QED, and SA, proposed objectives for ScaMARS to optimize included:  Calculated Partition Coefficient (cLogP), Number of Rotatable Bonds (nRotat), Fraction of scp$^3$ hybridized carbons (fCsp3), Molecular Weight (MW), Topological Polar Surface Area (TPSA), and Tanimoto similarity to the starting scaffold. \cite{QED,SynthAccess}

The default desirability function is a Derringer function for the additive mean of all normalized properties as follows:
\begin{equation}
(\sum_{i=1}^{n}d_{i}Y_{i})/n\label{eq:01}
\end{equation}
where $n$ is the number of properties and $d_{i}$ is the weight given to the normalized property $Y_{i}$. Each $Y_{i}$ was normalized with different user-defined functions according to the SwissADME property ranges and whether the value must be maximized or minimized. Most were a linear function between the maximum and minimum which defaults to 0 for values outside the given range. For this application, all property weights were kept at 1. The full equation for all ten properties would therefore be an average of raw values, normalized score, and conditionals such as QED, SA, and cLogP respectively, outlined in Table 1.

\begin{table}[ht]
\centering
\caption{Each property included in ScaMARS with its default scoring function. Numbers and trends follow filters outlined by SwissADME. The model accepts scores between 0 and 1 according to least and most desirable value respectively.}
    \begin{tabular}{c|ccccc}\hline
    Property                 & $Y_{i}$ Formula \\\hline
    QED                      & x  \\
    SA     & $(9-x)/10$   \\
    cLogP                    & \begin{tabular}[c]{@{}l@{}}$\begin{Bmatrix}
    (-0.7-x)/5.7 & ,-0.7<x<5\\ 
    0 & , otherwise
    \end{Bmatrix}$\end{tabular} \\
    TPSA                     & \begin{tabular}[c]{@{}l@{}}$\begin{Bmatrix}
    (130-x)/110 & ,20<x<130\\ 
    0 & , otherwise
    \end{Bmatrix}$\end{tabular}   \\
    nRotat                   & \begin{tabular}[c]{@{}l@{}}$\begin{Bmatrix}
    (10-x)/10 & ,x<10\\ 
    0 & , otherwise
    \end{Bmatrix}$\end{tabular}    \\
    MW                   & \begin{tabular}[c]{@{}l@{}}$\begin{Bmatrix}
    1 & ,150<x<500\\ 
    0 & , otherwise
    \end{Bmatrix}$\end{tabular}    \\
    fCSP3   & \begin{tabular}[c]{@{}l@{}}$\begin{Bmatrix}
    x & ,0.25<x\\  
    0 & , otherwise
    \end{Bmatrix}$\end{tabular} \\
    GSK3$\beta$ & x  \\
    JNK3    & x \\ 
    Tanimoto    & x \\ \hline
    \end{tabular}
\label{objtable}
\end{table}

Optimizing for a greater number of properties implies more balanced molecules, but increasing the number of properties in any additive mean has the unintended effect of decreasing the influence each has on the score. The alternative to this is a geometric mean. Every property has a strong effect on the score through multiplication and, if any property reaches zero, the entire score becomes zero and the molecule is ignored. This is closer to reality as well, as a toxic molecule will not be considered during drug development regardless of how desirable its other properties. Therefore, the geometric mean was implemented as an alternative:
\begin{equation}
(\prod_{i=1}^{n}d_{i}Y_{i})^{1/n}\label{eq:03}
\end{equation}
To mitigate unintended effects on the model, a hybrid of both additive and geometric was added where the model resorts to comparing the additive mean when no molecules in the new generation show improvement in geometric mean.
Linear normalization used a form already present in MARS, $\frac{max-x}{max-min}$, which would invert the property for minimization. Those that instead needed maximization were already on a scale of 0 to 1. Values beyond the extremes caused $Y_{i}$ to be set at 0, as that would imply toxicity or ineffectiveness.

\subsection*{Scaffold}

ScaMARS was run using the S-Adenosyl core as the scaffold for the results shown, as adenosine derivatives are of known biological importance \cite{adenosine_synthesis,kuroda1978physiological}. Prior runs using methyl or furan with varying numbers of properties did not significantly differ in runtime. Added flexibility and edits provide ScaMARS the choice to return to the scaffold if further proposals produce invalid molecules. When this occurs, there is a 50\% chance the path will propose the original scaffold instead of a modification. While the scaffold proposal must still be accepted through the MCMC sampler, acceptance is more likely to occur if the path is at a score closer to the original scaffold (little to no optimization in properties) or early in the run (higher annealing temperature). In the event more than one scaffold is input at the start, there is still a 50\% chance to return to a scaffold, after which the choice of specific scaffold is made at random.

\subsection*{Computation}

All ScaMARS trials were run for 5 hours on an RTX 2080 Ti GPU for 600 steps, each with a generation size of 1,000 molecules. Only the molecules produced at the final step were chosen for comparison with MolGCT, corresponding to the optima once the annealing temperature reached zero. MolGCT was trained using the same RTX 2080 Ti GPU as well for 48 hours for a total of 9 epochs. 2,000 molecules were generated using the trained molGCT with inputs logP=0.05 TPSA=20 QED=0.9, but only 890 remained valid and unique. Inputs were chosen to maximize properties while confined to the recommended ranges Kim et al. \cite{molGCT} set (those being 0.03-4.97 LogP, 17.92-112.83 TPSA, 0.58-0.95 QED). PaCMAP visualizations were created using the pacmap \cite{PaCMAP}, seaborn \cite{seaborn}, and RDKit \cite{RDKit} packages, with 2,048-bit Daylight fingerprints of each molecule as the high-dimensional space.

\section{Results \& Discussion}\label{sec3}

\subsection{Function Adaptations}

\begin{figure}[ht]
\begin{center}
\includegraphics[width=0.5\textwidth]{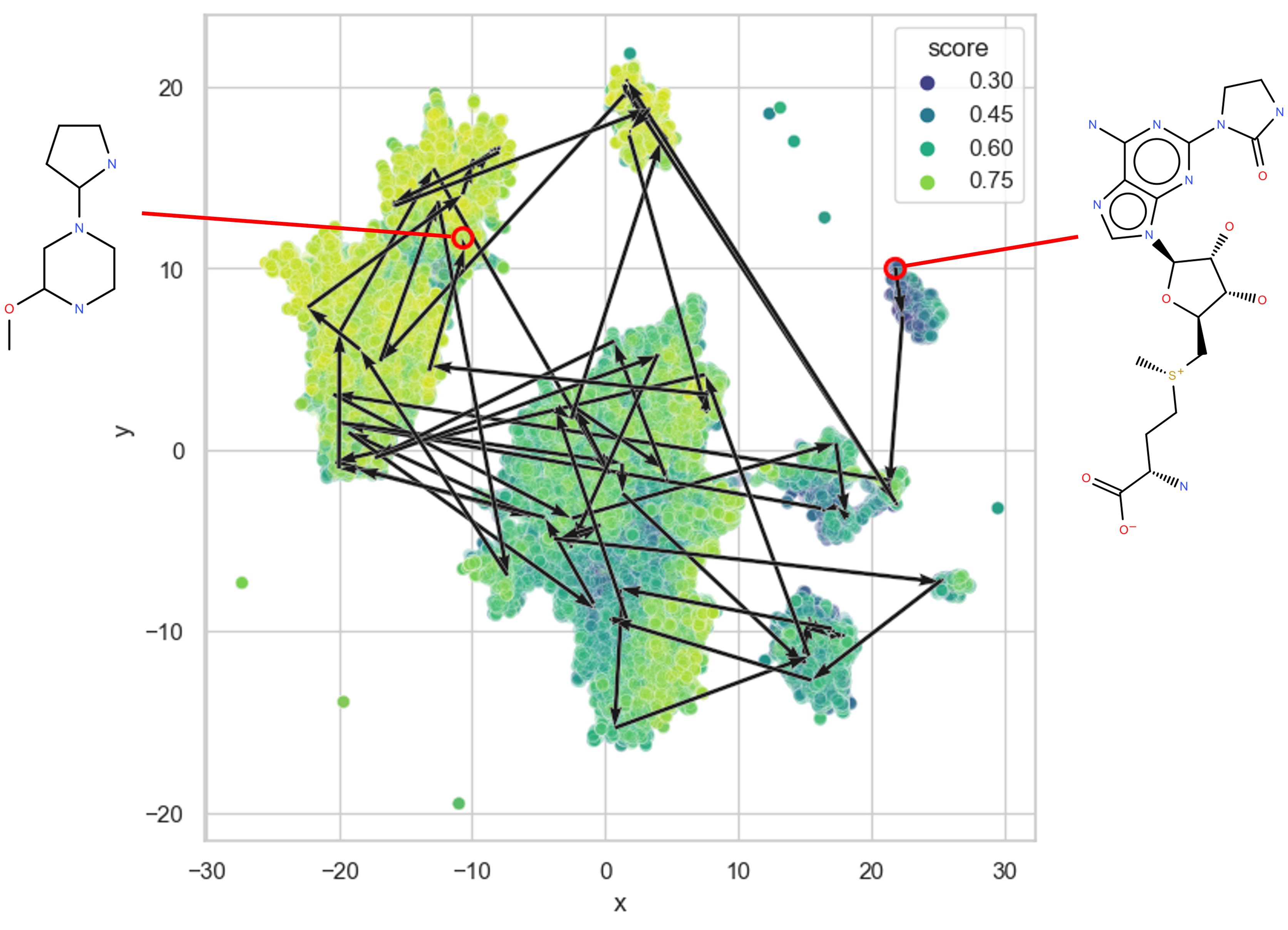}
\end{center}
\caption{Two embedding dimensions of a Pairwise Controlled Manifold Approximation (PaCMAP) using all 134,588 unique molecules produced during a ScaMARS run. 6 features were optimized (QED, TPSA, cLogP, nRotat, fCsp3, SA) starting from S-Adenosyl methionine (SAM) circled with red and illustrated on the right. Arrows track each modification this molecule underwent at each generation. Circled with red and on the left is the final molecule that was generated for one of the 1,000 paths.}\label{PaCMAP}
\end{figure}

Since ScaMARS is only capable of calculating MPNN loss for improved molecules, the use of a geometric mean in the objective function rendered the model nonfunctional. The geometric mean exclusively resulted in a score of 0 which prevented the model from learning and improving. The hybrid solution was intended to bridge the gap between allowing the model to learn and rejecting molecules more strictly, but it instead must resort to the additive method the majority of the time and increased computation time for each step. We did not observe an increase in ability to reach higher scores or raise scores quicker. The geometric mean is better suited for filtering final solutions, scoring the output molecules in post-processing more strictly. Nonetheless, the desirability function still serves as an efficient way of lowering the dimensionality of many properties. From 2 properties to 7, ScaMARS still completed 600 steps in under six hours. We clustered all molecules generated during the 6 property run using PaCMAP as shown in figure \ref{PaCMAP}.

\subsection{Comparison to molGCT}

Conditional models are limited by the training data, and generated molecules will be less intuitive. While it is beneficial to have control over exact property values, an optimization model can run quicker, show why a given structure was created, and propose molecules in different regions of chemical space according to a starting scaffold. According to metrics outlined by Xie et al. \cite{MARS} to compare MARS against other optimization models, molGCT performs similarly but lacks the ability to generate molecules that pass the SwissADME checks for Success Rate. In total, ScaMARS has a Diversity (Div) of 84.6\% while molGCT has 85.0\%. ScaMARS Success Rate (SR) totals 99.5\% while molGCT totals 52\%. The difference in SR is a result of TPSA values in molGCT molecules following a normal distribution between -1 and 2, potentially due to insufficient training or the model's inability to balance TPSA with other properties.

\subsection{Redundant Vocabulary}

Screening the 1,000 fragment vocabulary extracted from ChEMBL also revealed 164 redundant fragments, which limits efficient exploration of the chemical space and biased the model towards the repeated, but more drug-like, fragments. This is due to the model implicitly storing the connectivity of the fragment in the order of atoms in non-canon SMILES (ie. fragment OC (methanol) will attach to a scaffold C at its first atom, oxygen, making COC (methoxymethane) and never OCC (ethanol)). Fragments will, therefore, only use the valid and drug-like connections seen in the ChEMBL vocabulary, benefiting desirability at the cost of limiting solution space.

\subsection{Chemical Space}

\begin{figure}[ht]
\begin{center}
\includegraphics[width=0.5\textwidth]{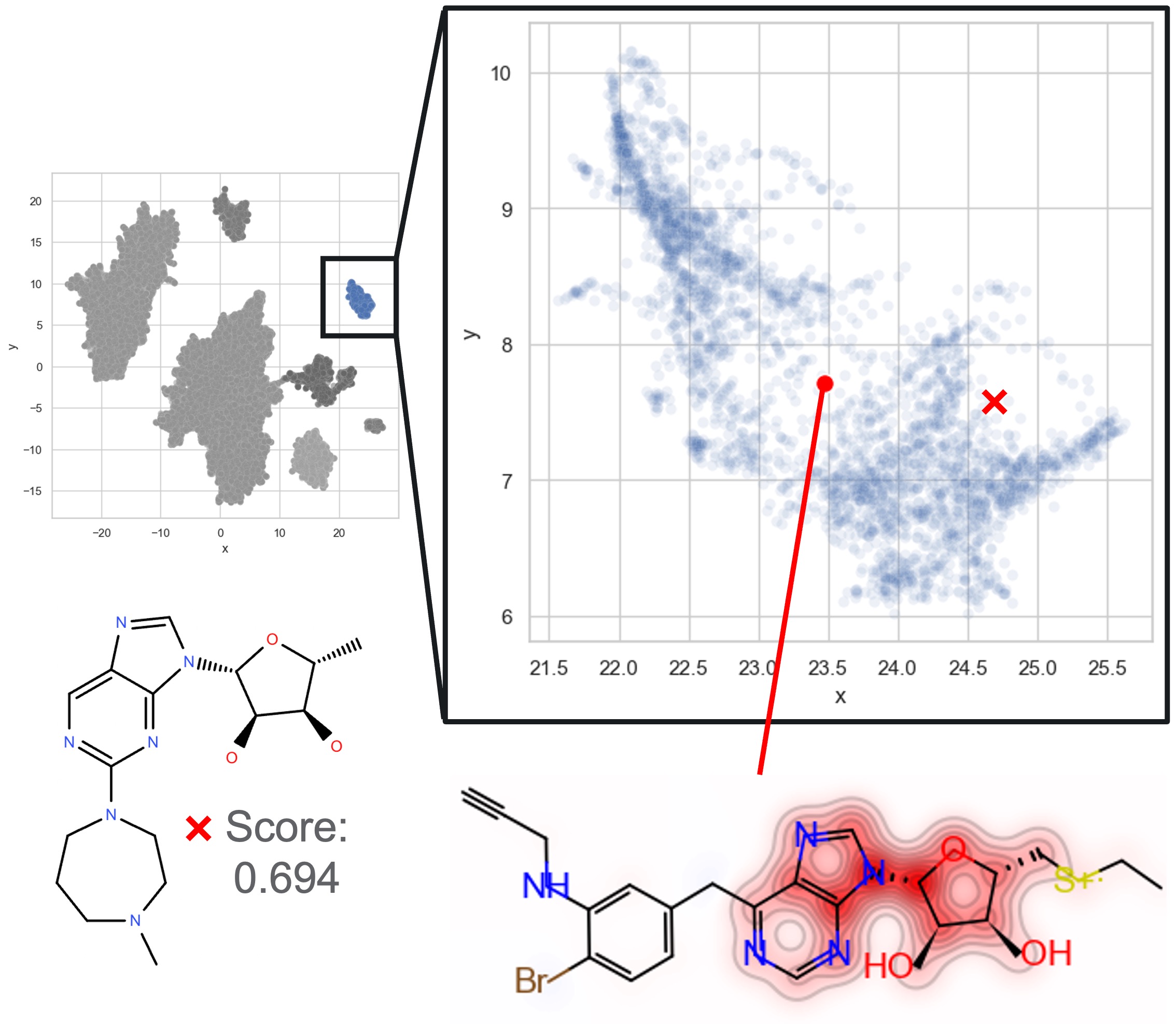}
\end{center}
\caption{Closer look and analysis of the 1,852 compound cluster housing the original starting SAM scaffold. The red line follows the medoid of the cluster to its point on the graph. Heatmap shading on the medoid compound correspond to each atom's contribution to the compound's average average Tanimoto similarity towards the rest of the cluster, with darker red being a greater loss to similarity through the removal of that atom. Labelled with a red 'X' on the graph and visualized to the left of the medoid is the highest-scoring compound in this cluster with a score of 0.694. 
}\label{medoid}
\end{figure}

\begin{figure}[ht]
\begin{center}
\includegraphics[width=0.8\textwidth]{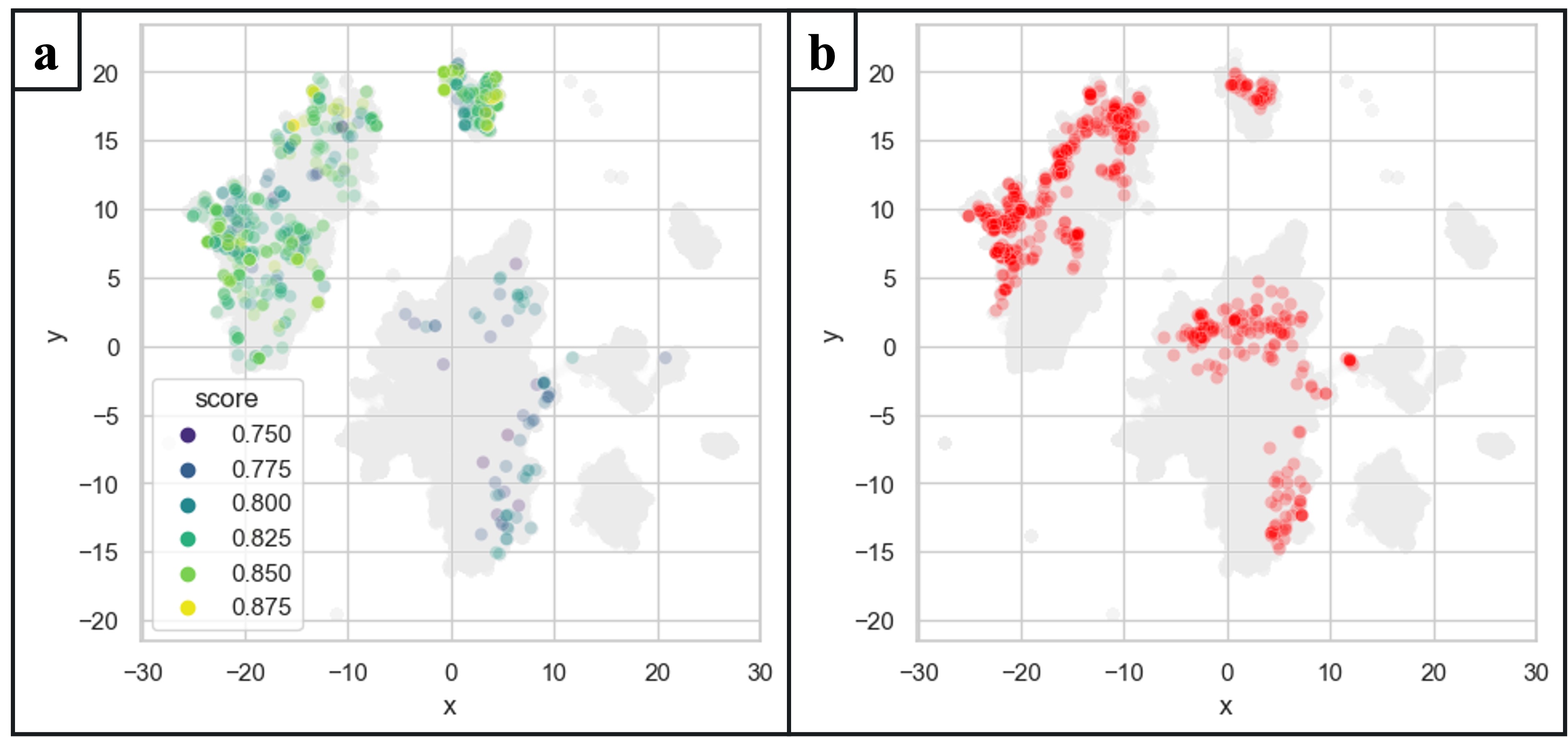}
\end{center}
\caption{PaCMAP of final molecules produced by either model embedded onto the previously calculated 134,588 molecule, 2048-dimensional space. (a) ScaMARS molecules colored by score. (b) molGCT molecules.}\label{PaCMAP_Split}
\end{figure}

To better explain the ScaMARS solutions and why certain clusters formed on the PaCMAP embedding, the medoid of each cluster was treated as a representative member. This finds the molecule in the cluster with the smallest average embedding distance to all other members of the group. Alternatively, centroids could be calculated but would include non-boolean fingerprint values or, restricting it through a step function, invalid structures. The medoids then had each atom weighted by its contribution towards the molecule's similarity to the rest of the cluster by removing one at a time and measuring the new average Tanimoto score. The result for the 1,852 compound cluster the initial SAM is shown as a heatmap over the medoid in Figure \ref{medoid}. The most common substructure for that cluster was the S-Adenosyl core originally present in SAM with the side-chains of the adenine rings commonly replaced. A SMARTS query could not find this substrucutre, its adenine half, nor its ribofuranose-like half present in any other cluster. This not only illustrates why this cluster forms but provides a novel observation of the original scaffold: the model was more likely to alter atoms attached to the core rather than the core itself based on a perceived score decrease when doing so. Otherwise, it would be possible to see accepted molecules elsewhere in the space with one half (such as purine) while the other is modified or removed. Figure \ref{medoid} also included the highest-scoring compound from that cluster. Its comparatively low score compared to final candidates is due to all molecules in this cluster struggling to increase SA and TPSA, possibly linked to the core scaffold they all share and reason for the model to quickly diverge from this substructure.

\subsection{Scaffold Choice}

The flexibility ScaMARS provides in the choice of starting scaffold, properties, and fragment vocabulary are essential to drug design and hindered by the focus on generating a diverse library. Paths will quickly put distance between the generated molecule and scaffolds in order to satisfy the novelty requirements. Figure \ref{PaCMAP} illustrates this when the path leaves the cluster around the initial SAM scaffold after two steps, consistent with the majority of paths reaching below 0.5 similarity to SAM within 25 generations. Instead of favoring the cluster around the given scaffold, the larger groups (indicative of where paths most frequently generated compounds that MCMC accepted) are more distant and separated. This results in a final molecule structurally unrelated to the original.

\begin{figure}[ht]
\begin{center}
\includegraphics[width=0.7\textwidth]{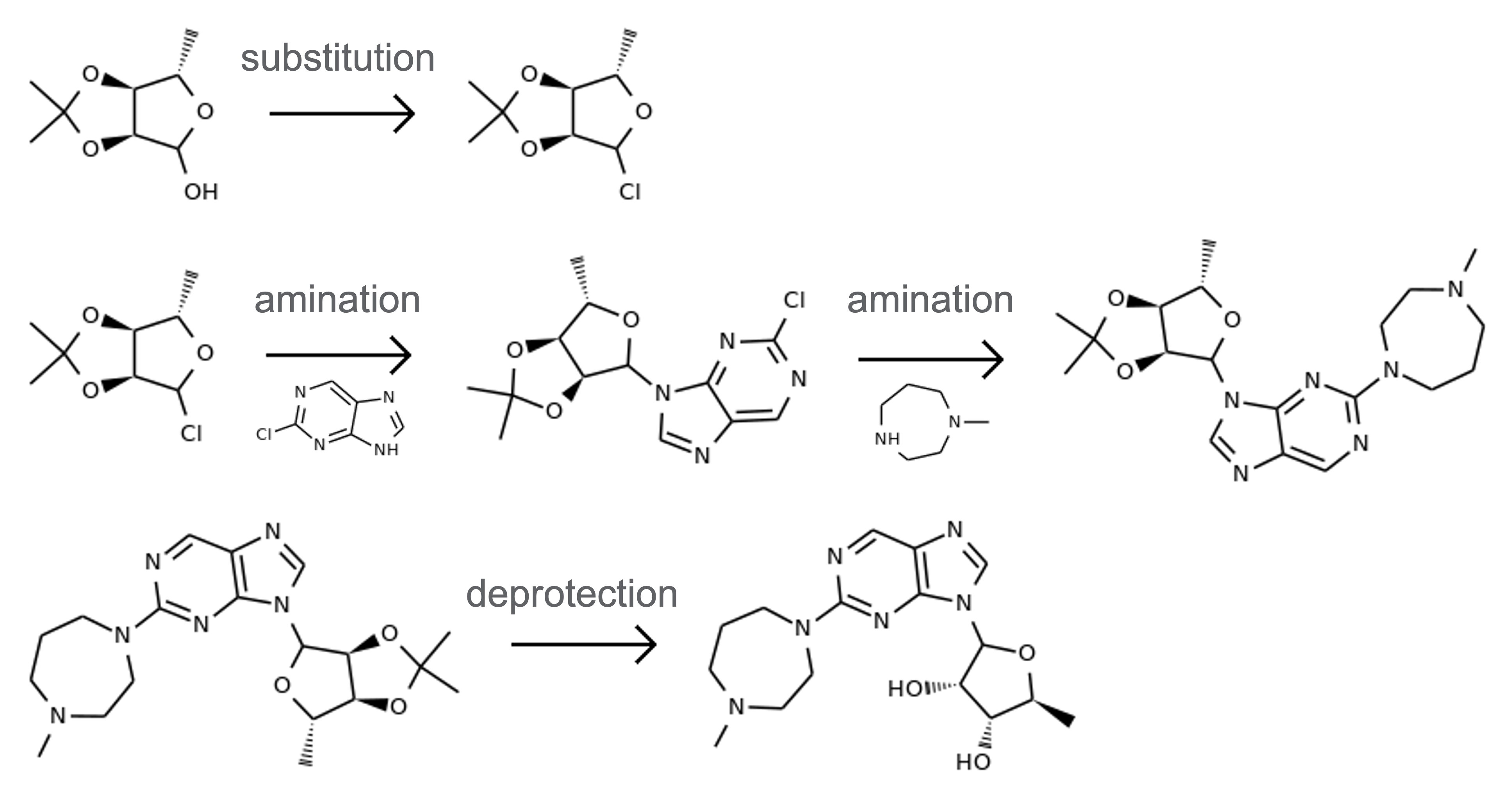}
\end{center}
\caption{Possible synthetic route proposed by ASKCOS for the highest-scoring SAM analogue generated through ScaMARS.}\label{synthesis}
\end{figure}

Predicting the synthesis of the highest scoring SAM cluster molecule through ASKCOS \cite{ASKCOS} reveals a path similar to known synthesis of adenosine derivatives. \cite{adenosine_synthesis} Depicted in Figure \ref{synthesis}, the three purchasable components (purine, 5-deoxy ribofuranose, and homopiperazine) are joined through aminations. Highest-scoring molecules without the adenosine scaffold were all structurally and synthetically similar, being chains of piperazines and homopiperazines linked either N to N or N-C-N. Though these are visually simple, ASKCOS could not predict a feasible synthetic route. Instead, it suggested reductions from larger molecules of which it is a substructure, but querying PubChem or MCULE returned no such purchasable compounds.

\subsection{Structural Trends}

Comparing the distributions in Figure \ref{PaCMAP_Split}, both models produce molecules in similar regions of greater MPO score, though each still has bias. For example, a greater amount of molGCT compounds are present around the center than ScaMARS. The center cluster's medoid indicates it is a region favoring the inclusion of multiple linked aromatic and aliphatic rings. Conversely, the top left grouping scores higher on average and seems solely concerned with aliphatic, heterocyclic nitrogen/carbon rings. This relates back to the highest-scoring compound of the SAM cluster also incorporating an aliphatic, heterocyclic nitrogen/carbon ring onto the core, indicating this substructure alone seems to frequently contribute to higher scores.

\section{Conclusion}\label{sec4}

In this contribution, we present the ScaMARS model with a flexible architecture for multiparameter optimization. This includes prioritization of the initial scaffold, support for a greater number of properties, and variants to the desirability function. Comparison to a Transformer conditional model molGCT shows that ScaMARS remains better suited for the optimization of molecules for drug design and discovery. 99.5\% of final ScaMARS molecules meet the desired property thresholds compared to only 52\% of molGCT. Both models converge on a 85\% diversity and occupy similar regions of chemical space (aliphatic N heterocycles) even though ScaMARS began sampling around known ligand SAM and took 32 fewer hours of preparation. Clustering and analysis of ScaMARS molecules revealed what substructures are most important or avoided during the optimization process.

The resulting compounds from our approach, like other methods, are structurally distinct from known compounds. Current challenges in restricting similarity and keeping high-dimensional pareto optimization computationally feasible present areas of improvement. In the future, we could apply existing XAI techniques to the model to compare gains in explainability, remove redundancy in the fragment-based and SMILES approach (such as with JANUS's SELFIES), and restrict the model to explore only the chemical space immediate to the chosen scaffold for drug design.

\section*{Declarations}

\subsection*{Funding}

This research was supported by the I3T Investment, under the Laboratory Directed Research and Development (LDRD) Program at Pacific Northwest National Laboratory (PNNL). PNNL is a multi-program national laboratory operated for the U.S. Department of Energy (DOE) by Battelle Memorial Institute under Contract No. DE-AC05-76RL01830. The computational work was performed using PNNL's research computing at Pacific Northwest National Laboratory.

\subsection*{Conflict of Interest}

The authors declare that the research was conducted in the absence of any commercial or financial relationships that could be construed as a potential conflict of interest.

\subsection*{Data Availability}
ScaMARS code and data is awaiting approval before public release.

The current public CHEMBL release was the only external dataset, serving as a comparison to the outdated file in the original MARS repository but never used in model training. Code edits and all generated data compared in analysis can be found in supplemental information files.\\
CHEMBL Dataset; \url{https://www.ebi.ac.uk/chembl/}\\
Xie et al. 2021. MARS; \url{https://github.com/bytedance/markov-molecular-sampling}\\
Kim et al. 2021. MolGCT; \url{https://github.com/Hyunseung-Kim/molGCT}

\subsection*{Author Contributions}

N.K. and A.K. designed the project. A.D.M. and A.K. analyzed the data. A.K. ran models, wrote the draft, and prepared figures. All authors contributed to revising and approving the final manuscript.

\bibliographystyle{unsrtnat}
\bibliography{references}  






\end{document}